\documentclass[a4j]{article}
\usepackage[dvips]{graphicx}

\begin{document}

\title{Evolution of Multispecificity in an Immune Network}
\author{
  Kouji Harada
    and 
      Takashi Ikegami 
        \vspace*{0.2cm}\\
          Institute of Physics, The Graduate School of Arts and Sciences,\\
            University of Tokyo,\\
              3-8-1, Komaba, Meguro-ku, Tokyo 153, Japan \\
               harada@sacral.c.u-tokyo.ac.jp \\
               ikeg@sacral.c.u-tokyo.ac.jp  \\
               }
              
\newenvironment{indention}[1]{\par
 \addtolength{\leftskip}{#1}
  \begingroup}{\endgroup\par}
   \maketitle
              
\begin{abstract}

  Divergence in antigen response of the immune network is discussed,
 based on shape-space modelling. The present model extends the
 shape-space model by introducing the evolution of specificity of
 idiotypes.  When the amount of external antigen increases, stability
 of the immune network changes and the network responds to the
 antigen. It is shown that specific and non-specific responses emerge
 as a function of antigen levels. A specific response is observed with
 a fixed point attractor, and a non-specific response is observed with
 a long-lived chaotic transient state of the lymphocyte population
 dynamics. The network topology also changes between these two states. 
 The relevance of such a long-lived transient state is discussed with
 respect to immune function.

\end{abstract}

\section{Introduction}

The `lock and key' concept has been central to understanding the
specificity of biochemical molecular interactions, from
enzyme-substrate relationships to antigen-antibody matchings. However,
it has gradually been realized that such a `lock and key' concept is
not strictly valid, particularly in immune systems
\cite{Ghosh}. Antigen-antibody interactions are found to be plastic or `
multispecific' rather than fixed or single-specific.  Namely,
antibodies inherently have a flexible recognition capacity. Kearney et
al.\cite{Kearney} have confirmed experimentally the existence of such
ambiguity of recognition in the antibody binding site of immature B
cells. It is generally believed that development from ambiguous to
specific recognition is caused by somatic hypermutations
\cite{wedemayer}. We here propose a new dynamics of specificity
evolutions based on Jerne's network hypothesis
\cite{jerne}. Our model is characterized by a meta-dynamics of
idiotype specificity on shape-space \cite{siegel}. We show here that
specific and non-specific responses to an antigen are governed
dynamically by a fixed point attractor and a chaotic long-lived
transient state of an immune network, respectively. The relevance of
such a long-lived transient state is discussed with respect to immune
function.

\section{Modeling with a Meta-dynamics of Specificity}

   We first introduce the standard idiotypic network model. Each
idiotype is characterized by a pair of surface sites, called the
idiotope and the paratope. If the idiotope site of a lymph cell is
bounded by paratopes of other lymph cells, the recognized lymph cells
become inactivated, whereas the recognizing cells become
activated. Thus the growth dynamics of clone size $x^n_{k,j}$ of an
idiotype of paratope $k$ and idiotope $j$ is given as,

\begin{equation}
  x^{n+1}_{k,j} = x^n_{k,j} + \sum_p \sum_q b_{q,k}x^n_{p,q}x^n_{k,j} 
 - \alpha \sum_p \sum_q b_{j,p}x^n_{k,j}x^n_{p,q} 
-d x^n_{k,j} + s,
\end{equation} 

The idiotope-paratope interaction $ b_{i,j} $ is assumed to have an
exponential form: $\frac{1}{\sigma} e^{\frac{-|i-j|}{\sigma}}$. We
characterize the ambiguity of the antigen-antibody by the deviation
parameter $\sigma$. The proposed meta-dynamics controls this
parameter. First, as a simple example, we quantize $\sigma$ by the
power of 2: $\sigma_{m} = 2^{M-m}$. The maximum specificity is given
by $m =M$.

Now each idiotype is characterized by three variables: idiotope $k$,
paratope $j$, and the specificity $m$. We thus describe the evolution
of specificity as follows:

\begin{equation}
  x^{n+1}_{k,j,m} = (1-\mu')x^{n}_{k,j,m} + 
\mu'/2 \sum_{m'= m-1,m+1} x^{n}_{k,j,m'} + s_{m=1},
\end{equation} 

where $\mu'$ is the mutation rate of specificity. Here the source term
$s_{m=1}$ is added for the least specific antibody. This dependency
reflects the fact that the premature B-cells are believed to have
lower specificities.

By combining these equations, we establish the complete clone growth
dynamics with mutations among idiotypes and the evolution of the
specificities.

In our model, there are five different types of idiotopes and of
paratopes, so that there are 25 different idiotypes, with $M=5$
different levels of specificity. The rest of the system parameters
(i.e. $\mu'=0.3$, $s=1.0$ , $d=0.1$, and $\alpha=2.0$) are selected so
that the size of each clone never diverges.

The following results (especially, the natural tolerance at high
amount of antigen) are confirmed not to depend on the values of the
mutation rate $\mu'$ and the source $s$. The dependency of system size
is still unclear.

\section{Dynamical Natures of the Network}

 We pay most attention to how the idiotype network responds to
persistent antigenic stimulations. A static antigen with a binding
site $k$ is introduced by adding the constant term $
+A_kb_{k,j}x_{i,j,m}$ to the above equation.  Estimating the mean
network specificity $Sp_k$ by averaging the specificity of all
idiotypes bearing paratope type {\sl k}, we study the antigenic effect
on the network dynamics.

An antigen of type 4 is used as an example, but the following result
does not depend on the selected antigen type. Because we adopt the
periodic boundary condition for the shape-space, each idiotype is
equivalent within a network.

\begin{figure}[t]
\begin{center}
\includegraphics[height=5cm,width=7.5cm,clip]{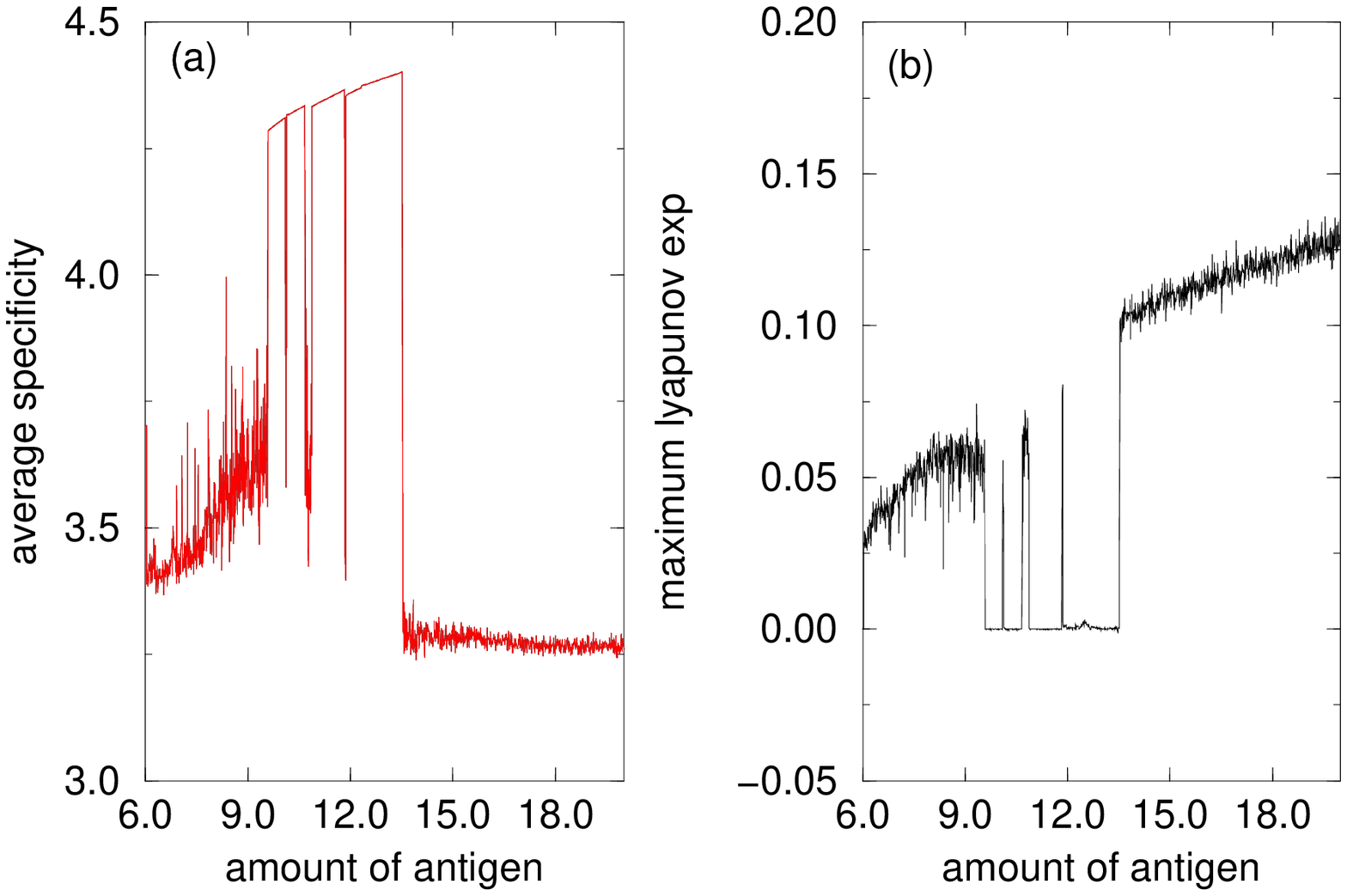}
\includegraphics[height=5cm,width=4.5cm,clip]{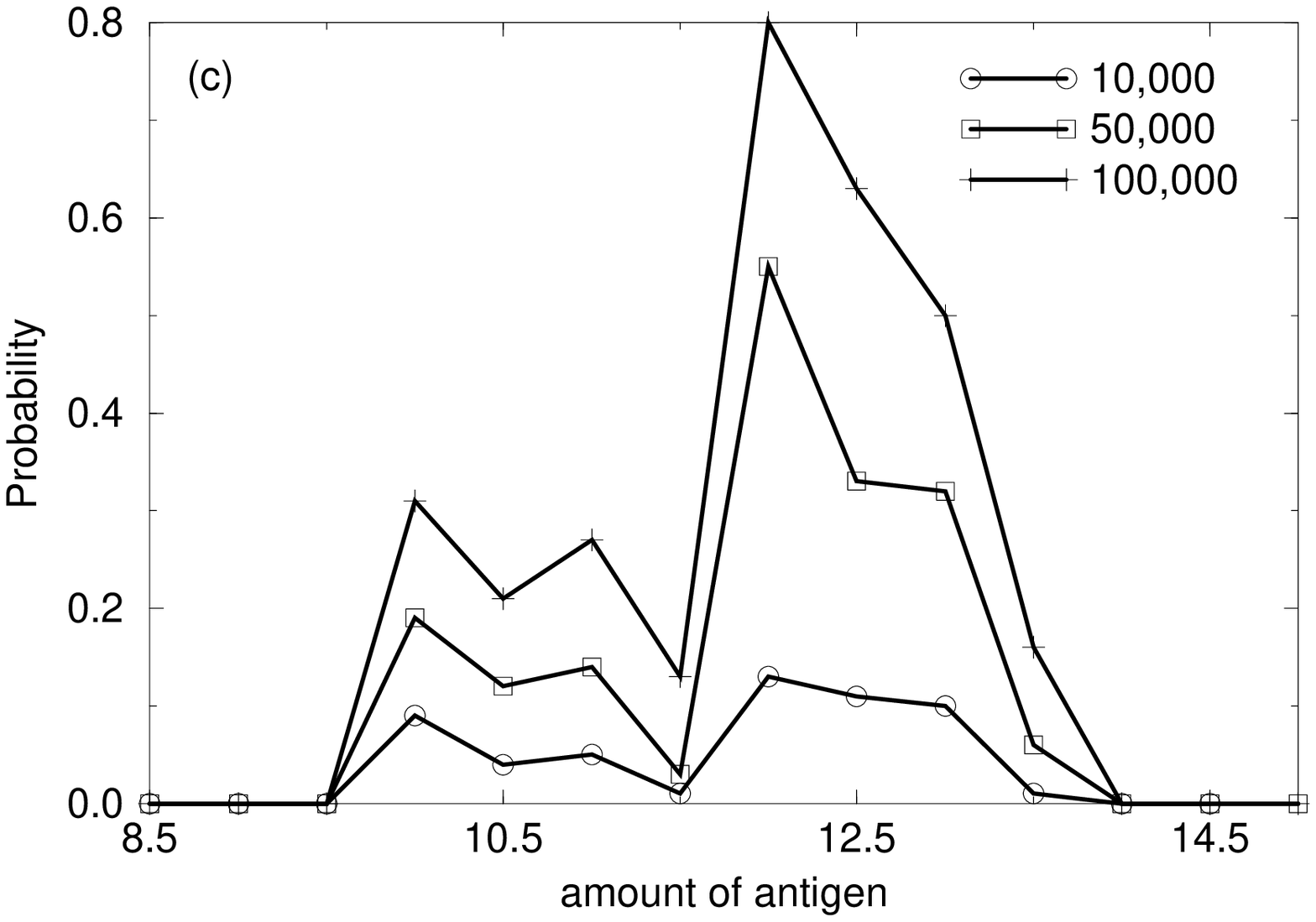}
\caption{The average network specificity and the maximum Lyapunov 
exponents are plotted against the antigen level in Fig. 1(a) and (b)
respectively, and in Fig. 1(c), the stability of the type I attractor
is plotted against the antigen level. Stability is measured by the
proportion of idiotypes in the initial distribution that show a
transition from a type I to a type II state before a given time
step. The time intervals used are 10,000, 50,000, and 100,000
steps. The number of initial distribution sets is 100.}
\label{sp_lyap}
\label{prob}
\end{center}
\end{figure}

 We show a plot of the averaged specificity ($\bar{Sp_4}$) and the
maximum Lyapunov exponent under the antigen stimulations over $10^4$
steps (see Fig. \ref{sp_lyap}(a), (b)).

In Fig.\ref{sp_lyap}(a), as expected, the network specificity
increases when we increase the amount of antigen. At about 9.5 units
of antigen, however, the specificity abruptly diverges to a high
value. We say that a specific response has occurred at this antigen
level. This specific response is observed until the antigen level
reaches 13.5 units. Beyond this critical value, the specific response
is no longer observed. Inversely, the specificity is sustained at the
lower values. This lower sustained response can be compared to natural
tolerance to the antigen.

On the other hand, by comparing Fig.\ref{sp_lyap}(a) with (b), when
the amount of the antigen is set between 9.5 and 13.5 units, we notice
that the lower specificity emerges with chaotic dynamics, and the
higher specificity emerges with a fixed point dynamics. We shall call
the former dynamics a type I attractor and the latter a type II
attractor.

However, the type I attractor is not a true attractor. It was found to
be a long-lived transient state referred to as a {\it super-transient
state}, which is a common phenomenon in high-dimensional dynamical
systems \cite{Kaneko}. In Fig.\ref{prob}(c), when the observation
period is extended, we observe a transition from type I to type II
attractor. There is no inverse-transition from the type II to the type
I attractor. The super-transient states are highly dependent on the
antigen level. For example, when the antigen level is 11.5 units in
Fig.\ref{prob}(c), the transition probability from type I to type II
is still less than 12 percent. In such cases, it behaves as an
attractor in a practical sense.

From a practical viewpoint, response time is also worth noticing. If
we say that the relevant time scale for the immune response should be
less than 10,000 time steps, in a practical sense there is no specific
response even at higher levels of antigen (see Fig.\ref{prob}(c)). Our
results suggest that a certain level of antigen causes the
super-transient state to suppress fast immune responses under the
idiotype network.

\begin{figure}[t]
\begin{center} 
\includegraphics[height=5cm,width=9cm,clip]{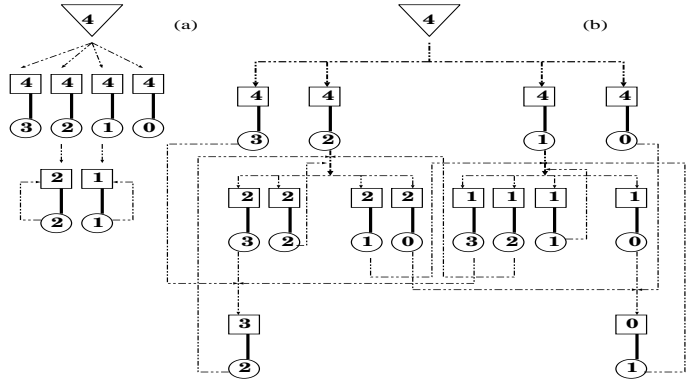}
\caption{
Topology of an idiotype network of type II and I attractors is shown
in Fig.2 (a) and (b), respectively, for an antigen level of 12 units.
Only idiotypes that have a population of 1.25 on average of over
10,000 time steps are depicted. Each idiotype in this figure is
represented by a pair of symbols, square and circle. The square with
the numeral inside denotes the paratope, whereas the circle with the
numeral inside denotes the idiotope. The triangle with the number
inside represents the injected antigen type. A stimulation wave from
idiotope to paratope is shown by a dotted line.}
\label{topology}
\end{center}   
\end{figure}

Besides the response time, much attention has been paid in the field
of theoretical immunology to topological changes of the network
\cite{bersini,calenbuhr,vincent}. Here we argue that the transition
from type I (unspecific) to type II (specific) causes a simultaneous
change of network topology. The network topology of each of these two
states is shown in Fig.\ref{topology}.

As we see from the figure, a chaotic super-transient state of type I
has a more complex network than does type II. Inversely, higher
specificity to the dosed antigen is maintained by a simpler network
structure. The maintenance of idiotypic diversity can be attributed to
chaotic dynamics.

By estimating the amount of specificity of all idiotypes in the type
I's distributed state and the type II's localized state respectively,
it is found that each idiotype in type I's distributed state has a low
specificity on the whole. Namely, each idiotype interacts weakly with
many idiotypes in order to have high connectivity. As a result, the
stimulation of the network by dosed antigen is distributed over the
network, not concentrated only on idiotypes bearing a binding site
(paratope with type 4) for the antigen. Thus, the immune response to
the antigen has a tendency to be suppressed. This result would support
Stewart's extrapolation that `The higher connectivity among
idiotypes, the greater the degree of tolerance' \cite{stewart}.

Recently, a chaotic oscillation was found experimentally in a natural
tolerant state. Subsequently, theoretical immunologists have tried to
establish `natural tolerance under chaotic dynamics' against a static
antigen \cite{bersini,calenbuhr}, though their simulation results show
difficulty establishing such a tolerance without assuming a special
network topology of an `odd-loop structure' and so-called `bell-shaped
function' as an activation function. We have shown how such a
tolerance can arise naturally under a chaotic dynamics, without these
assumptions, by adding an additional flexibility; i.e., meta-mutation
dynamics with specificity of idiotype. We have used a simple idiotypic
network model, and have not ventured to use the more complex '
bell-shaped function model' because of focusing on capabilities of the
meta-dynamics we introduced. Applying the meta-dynamics we introduced
here with the bell-shaped function model is left as a future problem.

\section{Concluding Remarks}

In this paper, we have expanded the possibilities of theoretical
immunology by introducing new meta-dynamics. We believe that the
immune response should be seen as having a more dynamic nature than
allowed by most current models \cite{Harada}, and that the specific
antigen-response and the dynamical percolation related to natural
tolerance are caused by the meta-dynamics controlling the degree of
specificity, as introduced here.

\end{document}